# Definition and Implications of the Digital Near-Death Experience: A Theoretical Essay on Preliminary Empiricism

*Research in Progress*

**Pedro Jácome de Moura Jr**
Universidade Federal da Paraíba, UFPB, Brazil
pjacome@sti.ufpb.br

## Abstract

People are increasingly adhering to social networking platforms (SNP), and such adhesion is often unreflective, which makes them alienate data, actions, and decisions to *tech giants*. This essay discusses what happens when, eventually, someone chooses to cancel their participation in a large SNP. This is a theoretical essay, whose narrative resembles a theoretical-empirical manuscript, grounded on the author's experience and his subjective perceptions regarding being out of the WhatsApp network (nowadays, the main SNP instance in the world). This study proposes a definition and implications of the supposedly new "digital near-death experience" concept, a metaphor for the well documented *near-death experience* (NDE). A research agenda is also proposed.

## Introduction

Recently I decided to cancel my WhatsApp account. It was a very easy decision and also an easy task. A day before, few friends and relatives were informed through my last message on the platform. There were no immediate repercussions in general, except some emojis and briefs "okay" in response, with just two messages questioning my motivations for such a decision. In fact, my reasoning was not precisely defined, but it went through the feeling of dependence of a large part of society in relation to a single information technology (IT) provider.

At that time I was aware of the negative effects of IT-related phenomena on individuals, which is not a novel subject. Addictive IT behavior, deceptive computer-mediated communication, IT interruptions, IT misuse, technostress, impulsive use of IT and physiological effects of IT usage are "potentially harmful" consequences of IT-related phenomena (Turel et al, 2017, p. 5648) and has been a serious matter of concern for information systems (IS) research (Boroon et al., 2021), as illustrated through a specific session held at the Hawaii International Conference on System Sciences (HICSS), promoted by the Association for Information Systems (AIS) in 2017.

While researchers have tried to figure out the role of the state in regulating social network platforms (SNP) (e.g. Van Dijck, 2020)—because counting on billions of users, SNP may concentrate unimaginable power and resources, and pose threats to democracy (e.g. Nemitz, 2018)—, the same has no correspondence when it comes to an individual's role in facing such a threat. I mean, most of the literature has discussed the political perspective of the problem, but this is not the only (or even the main) way forward. The path of the freethinker individual, who makes decisions and takes the consequences, must also be followed. It should emphasize the role of the agency, not just the structure.

One explanation for the emphasis on structure is disbelief in the agency to become aware of a given negative reality and, collectively, self-organize to overcome it (Gerbaudo, 2017). However, it seems paradoxical to delegate to the state the solution for confronting super-powerful players, when such a state is already super-powerful. Emphasizing the structure—of both super-powerful SNP or state—implies alienation of agency, since in a Hobbesian perspective, alienation means transferring to another the discretion to decide and act for oneself.

In the case of the negative effects of IT-related phenomena, the literature shows that internet addiction (Kim & Kim, 2002), smartphone addiction (Lăzăroiu et al., 2020), and social media addiction (Sun & Zhang, 2021) to mention a few, have been studied as forms of alienation. Presumably, facing alienation requires a profound change in the individual's willingness to take responsibility for the consequences of one's decisions and actions.

Back to canceling my WhatsApp account, I see the negative effects of IT as consequence of alienation and I envision the agency's role in solving this alienation as a complementary approach to the exclusive focus on structural aspects. But what would be the agency's subjective perceptions of being outside the specific social network? How long could the agency put up with abstinence from use? And what are the implications for the agency's alienation from stopping using this service?

Through a brief experimentation (a preliminary one) I found some answers which will support the design of a more comprehensive research agenda. In the next sections I describe and discuss the experience of being outside the specific social network, my feelings and behavior, ending up with theoretical implications and a research agenda proposal.

## The experience

The method can be summarized as (1) informing few friends and relatives (regular users of the SNP) a day before (17th May 2021). The message content was: "Dear all, soon I will be out of WhatsApp. The main contact will be via email. Bye"; (2) canceling the account (18th May 2021); (3) uninstalling the app from the smartphone and computer (18th May 2021); and (4) taking notes of subjective perceptions regarding being out of the SNP (for a month). Table 1 contains the transcript of the notes. Data analysis and recursive access to literature for interpretation of findings were inspired by grounded theory, according to Glaser & Strauss (1967), and abductive explanation (Haig, 2005).

**Table 1. Subjective perceptions regarding being out of the SNP**

| Notes | Time elapsed* | Nature** |
|---|---|---|
| Euphoria, enthusiasm | Seconds | Feeling |
| Feeling of being irrelevant | Hours | Feeling |
| Feeling of being forgotten. It must be like this when you die | Hours | Feeling |
| Frequent consultation with the wife (who continued to use WhatsApp) about the status (health, mood) of those close to her/him | 1 day | Behavior |
| Frequent inbox checking | 1 day | Behavior |
| Need to be physically closer to the smartphone to answer any phone calls (which did not occur before) | 1 day | Behavior |
| Need to return unanswered phone calls, as they might be important (which did not occur before) | 1 day | Behavior |
| Increased frequency of sending emails to friends and family | 2 days | Behavior |
| Outdating on ordinary and immediate matters | 2 days | Feeling |
| Difficulty in acquiring, hiring or solving problems with service providers (the vast majority use exclusively WhatsApp for communication) | 1 week | Behavior |
| Third party strangeness for not using WhatsApp (I'm seen as a freak). Some even smile playfully when I say I don't use this SNP | 1 week | Behavior |
| Account creation in another SNP, supported by a non-profit foundation | 1 week | Behavior |

| Anxiety for friends and family to gather in the new SNP | 1 week | Feeling |
|---|---|---|
| Partial adhesion of a few, especially family members and very close friends, to the alternative platform, but even so, with much greater latency between sending and reading messages | 1 month | Behavior |
| Suspension of personal projects in groups, due to the resistance of others to adopt alternative (not WhatsApp) means of communication | 1 month | Behavior |

Notes: (*) After cancelling the account; (**) This classification took place only after consulting the literature, especially on near-death experiences. Further details are provided in the "discussion" section.

## Discussion

Cancelling the WhatsApp account, the immediate sensations aroused by, and the return to the use of a SNP resemble, to some extent, a near-death experience (NDE). NDE is an "imminent death" situation (Egger, 1896, p. 26) which does not culminate with the end of life. During such an experience, an abrupt change in mental state occurs, followed by a return to the ordinary state. Still, the experience is transformative in the sense that it leaves traces, consequences of a "very surprising and consequential (or emotionally arousing) event" (Brown & Kulik, 1977, p. 73). In my case, it is more appropriate to classify the experience as a NDE-like, "since the phenomenology [of a NDE-like] is similar to a classical NDE yet without an actual imminent risk of death" (Martial et al., 2021, p. 1).

During a NDE or NDE-like, the temporary mental state is characterized by changes in perception of time, past, present and future (cognitive dimension); perception of joy, peace and harmony (affective dimension); non-existence/void/fear, point of no return, dying, return (border dimension); and perception of unearthly, mystical and religious figures (transcendental dimension) (Greyson, 1983; Martial et al., 2020). To analyze the notes recorded (Table 1), Greyson's (1983) "cognitive" and "affective" dimensions and Martial's et al (2020) "border" dimension were associated with subjective perceptions regarding being out of the WhatsApp, as described in the column "Nature" in Table 1.

Convergences that can be observed between the literature and the empirical data in this study suggest that the metaphor "digital near-death experience" (DNDE or e-NDE) is suitable for characterizing the near-death experience in the digital spacetime. First, a NDE implies a near-go and a return to ordinary life and in the DNDE/e-NDE case there is also a near-go (as in NDE) as there is an "emotionally arousing" event and, at the same time, the return to "life" through alternative IT resources. Second, once the negative effects of IT-related phenomena were identified, my immediate "treatment" was limited to discontinuing use but, as Vaghefi & Qahri-Saremi (2017) stated, abstinence ends up causing loss of volitional control, which sends the individual back to the repetitive pattern characteristic of addiction. That's what happened to me days after canceling my WhatsApp account. I started using another SNP.

Such a chain of events reinforces the adoption of NDE-like as a metaphor for the digital situation. It's digital near-death because I am back to using SNP again, just after the experience. It follows that the new DNDE/e-NDE concept can be described as "a mental state of imminent non-existence in a specific digital space-time". In this definition, "mental state" declares the transience of the experience; "imminence of non-existence" declares that this is a close experience, with great chances of coming true; and the mention of a "specific digital space-time" characterizes the technological support in which the experience takes place.

## Implications and research agenda

With billions of users, platforms like WhatsApp concentrate power and unimaginable resources. An example of such power is the dependence of a large part of society on a single service provider. Recent episodes of WhatsApp unavailability for a few hours and the impact of these events on users illustrate the dependency I am referring to (e.g. BBC, 2021; Guardian, 2020).

In this short essay (looking like a theoretical-empirical manuscript) I describe my own experience in reaction to this dependence. My subjective perceptions of being outside the specific SNP moved between enthusiasm, anxiety and frustration. There was no regret, but I didn't completely overcome the abstinence crisis and tried to bypass it through using another SNP. The facts of thinking about addiction and staying out of WhatsApp while I'm writing this report, illustrate that at least I reduced my level of alienation, a little.

Such reflections (answering the research questions) makes me question the current availability and sufficiency of theoretical support for the description, explanation and prediction of this class of phenomena in the literature. This essay suggests people need for contact and convergence to where (physically or virtually) everyone is. Why do people prefer/like to be where people are? An example is the empty pub versus the crowded pub. It doesn't matter the service (to some extent, of minimal quality), it matters to be among people. The same goes for a social network platform. It doesn't matter the negative effects of IT-related phenomena on individuals or the community, it matters to be among people. So, we need a socio-technical theory of density-attraction to shed light on this matter. As far as I know, such a theory does not yet exist, but it could well start from the immense contributions of Berger & Luckmann (1967) on socially constructed reality and of Giddens (1984) on the role of structure in shaping social behavior.

An initial set of assumptions for such a theory is (1) people need social support; (2) it is easier for people to be in evidence when sharing a space-time in which others (of interest) already are; (3) people are afraid of isolation; and (4) people are afraid of confronting themselves. Some initial hypotheses to be tested: (a) the density of participants in an SNP exerts a proportional force of attraction on those who do not yet participate; (b) the density of participants in an SNP influences the take-for-granted acceptance and use (alienation) by the participants; (c) the density of participants in an SNP negatively influences the decision to cancel participation; (d) the density of participants in a SNP positively influences the occurrence of "digital near-death experiences" in those who decide to cancel participation; (e) "digital near-death experiences" induce ($e_1$) anxiety, ($e_2$) frustration, ($e_3$) abstinence crises and ($e_4$) search for compensation, such as the return to addictive behavior; and (f) "digital near-death experiences" increase the valuation of free time.

The near-death experience (NDE; Greyson, 1983) and the near-death experience content (NDE-C; Martial et al., 2020) scales have been used for measurement of NDE and NDE-like self-reported occurrences by people after an episode of near-death. It is likely, however, that such scales are not sufficient to measure the digital expression of NDE. In line with this, the continuity of this research should include the validation of the new concept, and means of measurement and discrimination between NDE and DNDE/e-NDE. The development of a specific instrument for DNDE/e-NDE measurement should consider perceptions of who leaves and also of those who stay. To paraphrase several poets, "to die is to say goodbye". After all, who has been "abandoned", what do they perceive?

# References

BBC. (2021). Facebook, Whatsapp and Instagram back after outage. *BBC News*, available at https://www.bbc.com/news/technology-58793174.

Berger, P. & Luckmann, T. (1967). *The social construction of reality. A treatise in the sociology of knowledge*. London: Penguin Books.

Boroon, L., Abedin, B., & Erfani, E. (2021). The dark side of using online social networks: A review of individuals' negative experiences. *Journal of Global Information Management (JGIM)*, *29*(6), 1-21.

Brown, R., & Kulik, J. (1977). Flashbulb memories. *Cognition*, *5*(1), 73–99.

Egger, V. (1896). Le moi des mourants. *Revue Philosophique de la France et de l'Etranger*, 41, 26-38.

Gerbaudo, P. (2017). Social media teams as digital vanguards: The question of leadership in the management of key Facebook and Twitter accounts of Occupy Wall Street, Indignados and UK Uncut. *Information, Communication & Society*, *20*(2), 185-202.

Giddens, A. (1984). *The constitution of society: Outline of the theory of structuration*. Berkeley and Los Angeles: University of California Press.

Glaser, B. G. & Strauss A. L. (1967). *The discovery of grounded theory: Strategies for qualitative research*. New Brunswick and London: Transaction Publishers.

Greyson, B. (1983). The near-death experience scale. *Journal of nervous and mental disease*, *171*(6), 369-375.

Haig, B.D. (2005). An abductive theory of scientific method. *Psychological Methods*, *10*(4), 371.

Kim, S. W., & Kim, R. D. (2002). A study of Internet addiction: Status, causes, and remedies-focusing on the alienation factor. *International Journal of Human Ecology*, *3*(1), 1-19.

Lăzăroiu, G., Kovacova, M., Siekelova, A., & Vrbka, J. (2020). Addictive behavior of problematic smartphone users: The relationship between depression, anxiety, and stress. *Review of Contemporary Philosophy*, *19*, 50-56.

Martial, C., Fontaine, G., Gosseries, O., Carhart-Harris, R., Timmermann, C., Laureys, S., & Cassol, H. (2021). Losing the self in near-death experiences: The experience of ego-dissolution. *Brain Sciences*, *11*(7), 929.

Martial, C., Simon, J., Puttaert, N., Gosseries, O., Charland-Verville, V., Nyssen, A. S., ... & Cassol, H. (2020). The near-death experience content (NDE-C) scale: Development and psychometric validation. *Consciousness and Cognition*, *86*, 103049.

Nemitz, P. (2018). Constitutional democracy and technology in the age of artificial intelligence. *Philosophical Transactions of the Royal Society A: Mathematical, Physical and Engineering Sciences*, *376*(2133), 20180089.

Sun, Y., & Zhang, Y. (2021). A review of theories and models applied in studies of social media addiction and implications for future research. *Addictive Behaviors*, *114*, 106699.

Guardian. (2020). Facebook, Instagram and WhatsApp hit by media messaging outage. *The Guardian*, available at https://www.theguardian.com/technology/2019/jul/03/instagram-whatsapp-facebook-media-files-outage.

Turel, O., Soror, A., & Steelman, Z. (2017). The dark side of information technology: Mini-track introduction. In *Proceedings of the 50th Hawaii International Conference on System Sciences (HICSS)* (pp. 5648-5649).

Vaghefi, I., & Qahri-Saremi, H. (2017). From IT addiction to discontinued use: A cognitive dissonance perspective. In *Proceedings of the 50th Hawaii International Conference on System Sciences (HICSS)* (pp. 5650-5659).

Van Dijck, J. (2020). Seeing the forest for the trees: Visualizing platformization and its governance. *New Media & Society*, 1461444820940293.